\documentclass[prl,twocolumn,showpacs,amsmath,amssymb,aps,superscriptaddress]{revtex4-1}

\usepackage{graphicx}
\usepackage{dcolumn}
\usepackage{bm}
\usepackage{txfonts}
\usepackage{amsmath}
\usepackage{epsf}
\usepackage{epsfig}
\usepackage{amssymb}
\usepackage{color}
\usepackage{multirow}
\usepackage{pst-3d}
\usepackage{rotating}

\begin{document}

\title{Quantum Synchronization of Two Ensembles of Atoms}

\author{Minghui Xu, D. A. Tieri, E. C. Fine, James K. Thompson, and M. J. Holland}

\affiliation{JILA, National Institute of Standards and Technology and
  Department of Physics, University of Colorado, Boulder, Colorado
  80309-0440, USA}

\date{\today}%

\begin{abstract}
  We propose a system for observing the correlated phase dynamics of
  two mesoscopic ensembles of atoms through their collective coupling
  to an optical cavity. We find a dynamical quantum phase transition
  induced by pump noise and cavity output-coupling. The spectral
  properties of the superradiant light emitted from the cavity show
  that at a critical pump rate the system undergoes a transition from
  the independent behavior of two disparate oscillators to the
  phase-locking that is the signature of quantum synchronization.
\end{abstract}


\maketitle

Synchronization is an emergent phenomenon that describes coupled
objects spontaneously phase-locking to a common frequency in spite of
differences in their natural frequencies~\cite{book1}.  It was
famously observed by Huygens, the seventeenth century clock maker, in
the antiphase synchronization of two maritime pendulum
clocks~\cite{Huygens}. Dynamical synchronization is now recognized as
ubiquitous behavior occurring in a broad range of physical, chemical,
biological, and mechanical engineering systems~\cite{book1, book2,
  book3}.

Theoretical treatments of this phenomenon are often based on the study
of phase models~\cite{kuramoto0,kuramoto}, and as such have been
applied to an abundant variety of classical systems, including the
collective blinking of fireflies, the beating of heart cells, and
audience clapping.  The concept can be readily extended to systems
with an intrinsic quantum mechanical origin such as nanomechanical
resonators~\cite{Cross04,Milburn12}, optomechanical
arrays~\cite{Marquardt11}, and Josephson
junctions~\cite{Jain84,Wiesenfeld96}. When the number of coupled
oscillators is large, it has been demonstrated that the onset of
classical synchronization is analogous to a thermodynamic phase
transition~\cite{Winfree67} and exhibits similar scaling
behavior~\cite{Oleg09}.

Recently, there has been increasing interest in exploring
manifestations in the quantum realm.  Small systems have been
considered, {\em e.g.}, one qubit~\cite{Zhirov06} and two
qubits~\cite{Zhirov09} coupled to a quantum dissipative driven
oscillator, two dissipative spins~\cite{Orth10}, two coupled
cavities~\cite{Tony12}, and two micromechanical
oscillators~\cite{mian12,Mari12}. Connections between quantum
entanglement and synchronization have been revealed in continuous
variable systems~\cite{Mari12}. It has been shown that quantum
synchronization may be achieved between two canonically conjugate
variables~\cite{Hriscu13}.  Since the phenomenon is inherently
non-equilibrium, all of these systems share the common property of
competition between coherent and incoherent driving and dissipative
forces.

In this paper, we propose a modern-day realization of the original
Huygens experiment~\cite{Huygens}. We consider the synchronization of
two active atomic clocks coupled to a common single-mode optical
cavity. It has been predicted that in the regime of steady-state
superradiance~\cite{Meiser09,Meiser101,Thompson12,Thompson121} a
neutral atom lattice clock could produce an ultracoherent optical
field with a quality factor (ratio of frequency to linewidth) that
approaches~$10^{18}$. We show that two such clocks may exhibit a
dynamical phase transition~\cite{Zoller10,Zoller11,Cirac12,Cirac13}
from two disparate oscillators to quantum phase-locked dynamics. The
onset of synchronization at a critical pump strength is signified by
an abruptly increased relative phase diffusion that diverges in the
thermodynamic limit.  Besides being of fundamental importance in
nonequilibrium quantum many-body physics, this work could have broad
implications for many practical applications of ultrastable lasers and
precision measurements~\cite{Meiser09}.

\begin{figure}[b]
  \centerline{\includegraphics[width=0.7\linewidth, angle=0]{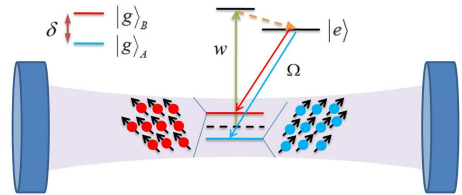}}
  \caption{\label{Fig1}(color online) Two ensembles of driven
    two-level atoms coupled to a single-mode cavity field. The atoms
    in ensemble $A$ are detuned above the cavity resonance (dashed
    line). Ensemble $B$ contains atoms detuned below the cavity
    resonance by an equivalent amount.}
\end{figure}

The general setup is shown schematically in Fig.~\ref{Fig1}. Two
ensembles, each containing $N$ two-level atoms with excited
state~$|e\rangle$ and ground state~$|g\rangle$, are collectively
coupled to a high-quality optical cavity.  The transition frequencies
of the atoms in ensembles $A$ and $B$ are detuned from the cavity
resonance by $\delta/2$ and $-\delta/2$ respectively. This could be
achieved by spatially separating the ensembles and applying an
inhomogeneous magnetic field to induce a differential Zeeman
shift. The atoms in both ensembles are pumped incoherently to the
excited state, as could be realized by driving a transition to a third
state that rapidly decays to
$|e\rangle$~\cite{Thompson12,Thompson121}.

This system is described by the Hamiltonian in the rotating frame of
the cavity field:
\begin{equation}
  \hat{H}=\frac{\hbar\delta} 2(\hat{J}_A^z
  -\hat{J}_B^z)
  +\frac{\hbar\Omega}{2}(\hat{a}^\dagger\hat{J}_A^-+\hat{J}_A^+\hat{a}
  +\hat{a}^\dagger\hat{J}_B^-+\hat{J}_B^+\hat{a})\,,
\end{equation}
where $\Omega$ is the atom-cavity coupling, and $\hat{a}$ and
$\hat{a}^\dagger$ are annihilation and creation operators for cavity
photons. Here
$\hat{J}_{A,B}^z=\frac{1}{2}\sum_{j=1}^N\hat{\sigma}_{(A,B)j}^z$ and
$\hat{J}_{A,B}^-=\sum_{j=1}^N\hat{\sigma}_{(A,B)j}^-$ are the
collective atomic spin operators, written in terms of the Pauli
operators for the two-level system $\hat{\sigma}_{(A,B)j}^z$ and
$\hat{\sigma}_{(A,B)j}^-=(\hat{\sigma}_{(A,B)j}^+)^\dagger$.

In addition to the coherent atom-cavity coupling, incoherent processes
are critical and include: the cavity intensity decay at rate $\kappa$,
the pump rate $w$, the free-space spontaneous emission rate $\gamma$,
and a background dephasing of the $|e\rangle$--$|g\rangle$
transition at rate $T_2^{-1}$. The total system is then described
using a master equation for the reduced density operator $\rho$:
\begin{eqnarray}
  \frac{d\rho}{dt}&=&
  \frac{1}{i\hbar}[\hat{H},\rho]+\kappa\mathcal{L}[\hat{a}]\,\rho
  +\sum_{\mathcal{T}=A,B}\sum_{j=1}^N\Bigl(
  \gamma_s\mathcal{L}[\hat{\sigma}_{\mathcal{T}j}^-]
  \nonumber\\ &&{}
  +w\mathcal{L}[\hat{\sigma}_{\mathcal{T}j}^+]+\frac{1}{2T_2}
  \mathcal{L}[\hat{\sigma}_{\mathcal{T}j}^z]\Bigr)\,\rho,
\end{eqnarray}
where $\mathcal{L}[\hat{O}]\,\rho=(2\hat{O}\rho
\hat{O}^\dagger-\hat{O}^\dagger \hat{O}\rho-\rho \hat{O}^\dagger
\hat{O})/2$ denotes the Lindblad superoperator.

The regime of steady-state superradiance is defined by the cavity
decay being much faster than all other incoherent
processes~\cite{Meiser09,Meiser101,Thompson12,Thompson121}. In this
regime, the cavity can be adiabatically eliminated~\cite{Meiser101},
resulting in a field that is slaved to the collective atomic dipole of
the two ensembles of atoms:
\begin{equation}\label{slave}
\hat{a}\simeq-\frac{i\Omega}{\kappa+i\delta}\hat{J}_A^-
-\frac{i\Omega}{\kappa-i\delta}\hat{J}_B^-.
\end{equation}
For small detuning on the scale of the cavity linewidth,
$\delta\ll\kappa$, Eq.~(\ref{slave}) reduces to
$\hat{a}\simeq-i\Omega\hat{J}^-/\kappa$, where $\hat{J}^-=
\hat{J}_{A}^-+\hat{J}_{B}^-$ is the total collective spin-lowering
operator. In this limit, the net effect of the cavity is to provide a
collective decay channel for the atoms, with rate
$\gamma_c=\Omega^2/\kappa$. This collective decay should be dominant
over other atomic decay processes~\cite{Meiser101}, {\em i.e.},
$N\gamma_c\gg\gamma_s, T_2^{-1}$, so that the time evolution is
effectively given by a superradiance master equation containing only
atoms:
\begin{equation}\label{master}
  \frac{d\rho}{dt}=\frac{\delta}{2i\hbar}[ J_A^z-J_B^z,\rho]
  +\gamma_c\mathcal{L}[\hat{J}^-]\,\rho
  +w\sum_{j=1}^N(\mathcal{L}[\hat{\sigma}_{Aj}^+]
    +\mathcal{L}[\hat{\sigma}_{Bj}^+])\,\rho.
\end{equation}

With this system we naturally provide the three necessary ingredients
for quantum synchronization; a controllable difference between the
oscillation frequencies of two mesoscopic ensembles, a dissipative
coupling generated by the emission of photons into the same cavity
mode, and a driving force produced by optical pumping.

The photons emitted by the cavity provide directly measurable
observables. Synchronization is evident in the properties of the
photon spectra.  In the case of two independent ensembles in the
unsynchronized phase, each ensemble radiates photons at its own
distinct transition frequency. This leads to two Lorentzian peaks that
are typically well-separated. In the synchronized phase, all of the
atoms radiate at a common central frequency resulting in a single
peak.

To solve this problem and find the steady state, we use a
semiclassical approximation that is applicable to large atom numbers.
Cumulants for the expectation values of system
operators~$\{\hat{\sigma}_{(A,B)j}^z,\hat{\sigma}_{(A,B)j}^{\pm}\}$
are expanded to second order~\cite{Meiser09,Meiser101}. All
expectation values are symmetric with respect to exchange of atoms
within each ensemble, {\em i.e.}\ $\langle\hat{\sigma}_{Bi}^+
\hat{\sigma}_{Bj}^-\rangle=\langle\hat{\sigma}_{B1}^+
\hat{\sigma}_{B2}^-\rangle$, for all $i\neq j$. Due to the U(1)
symmetry, $\langle\hat{\sigma}_{(A,B)j}^{\pm}\rangle=0$. Therefore,
all nonzero observables can be expressed in terms of
$\langle\hat{\sigma}_{(A,B)j}^z\rangle$,
$\langle\hat{\sigma}_{(A,B)i}^+ \hat{\sigma}_{(A,B)j}^-\rangle$, and
$\langle\hat{\sigma}_{(A,B)i}^z
\hat{\sigma}_{(A,B)j}^z\rangle$. Expectation values involving only one
ensemble are the same for both ensembles and for these cases we omit
the superfluous $A$,$B$ subscripts. The equations of motion can then
be found from Eq.~(\ref{master}):
\begin{eqnarray}\label{eqz}
  \frac{d}{dt}\langle\hat{\sigma}_{1}^z\rangle
  &&=-\gamma_c\left(\langle\hat{\sigma}_1^z\rangle+1\right)
  -w\left(\langle\hat{\sigma}_1^z\rangle-1\right)\nonumber\\
  &&{}-2\gamma_c (N-1)\langle\hat{\sigma}_1^+\hat{\sigma}_2^-\rangle
  -\gamma_c
  N\left(\langle\hat{\sigma}_{A1}^+\hat{\sigma}_{B1}^-\rangle+{\rm c.c}
  \right),\\
\label{eqx}
\frac{d}{dt}\langle\hat{\sigma}_1^+\hat{\sigma}_2^-\rangle
&&=-(w+\gamma_c)\langle\hat{\sigma}_1^+\hat{\sigma}_2^-\rangle+\frac{\gamma_c}{2}
\left(\langle\hat{\sigma}_1^z\hat{\sigma}_2^z\rangle
  +\langle\hat{\sigma}_1^z\rangle\right)\nonumber\\
&&{}+\gamma_c(N-2)\langle\hat{\sigma}_1^z\rangle
\langle\hat{\sigma}_1^+\hat{\sigma}_2^-\rangle\nonumber\\
&&{}+
\frac{\gamma_c}{2}N\langle\hat{\sigma}_1^z\rangle\left(
  \langle\hat{\sigma}_{A1}^+\hat{\sigma}_{B1}^-\rangle+{\rm c.c}\right),\\
\label{eqy}
\frac{d}{dt}\langle\hat{\sigma}_{A1}^+\hat{\sigma}_{B1}^-\rangle
&&=-(w+\gamma_c-i\delta)\langle\hat{\sigma}_{A1}^+\hat{\sigma}_{B1}^-\rangle
+\frac{\gamma_c}{2}\left(\langle\hat{\sigma}_{A1}^z\hat{\sigma}_{B1}^z\rangle
  +\langle\hat{\sigma}_{1}^z\rangle\right)\nonumber\\
&&{}+\gamma_c(N-1)\langle\hat{\sigma}_{1}^z\rangle
\left(\langle\hat{\sigma}_{A1}^+\hat{\sigma}_{B1}^-\rangle
+\langle\hat{\sigma}_{1}^+\hat{\sigma}_{2}^-\rangle\right),
\end{eqnarray}
describing population inversion, spin-spin coherence within each
ensemble, and correlation between ensembles, respectively. In deriving
Eq.~(\ref{eqx}) and (\ref{eqy}), we have dropped third order
cumulants~\cite{semi}. We also factorize
$\langle\hat{\sigma}_{(A,B)i}^z\hat{\sigma}_{(A,B)j}^z\rangle
\approx\langle\hat{\sigma}_{1}^z\rangle^2$, which we find to be valid
outside the regime of very weak pumping where a non-factorizable
subradiant dark state plays an important role~\cite{Meiser101}. After
making these approximations, Eq.~(\ref{eqz}) to (\ref{eqy}) form a
closed set of equations. The steady state is found by setting the time
derivatives to zero and the resulting algebraic equations can be
solved exactly. These solutions are the basis for the figures shown
below.
\begin{figure}[h]
  \centerline{\includegraphics[width=0.7\linewidth, angle=0]{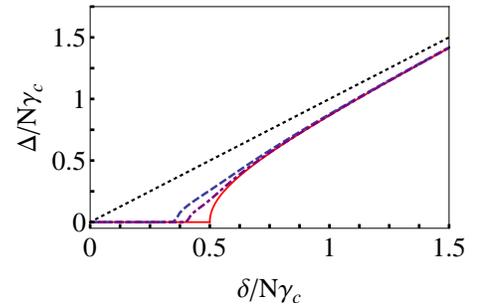}}
  \caption{\label{Fig2}(color online)~Steady-state relative phase
    precession for two ensembles as a function of detuning at
    $w=N\gamma_c/2$ for $N=100$~(blue dashed line), $N=500$~(purple
    dot dashed line) and $N=10^6$~(red solid line). The straight
    dotted line is $\delta=\Delta$.}
\end{figure}

In order to calculate the photon spectrum, we employ the quantum
regression theorem~\cite{qr} to obtain the two-time correlation
function of the light field,
$\langle\hat{a}^\dagger(\tau)\hat{a}(0)\rangle$, where time 0 denotes
an arbitrary time-origin in steady-state. In the limit
$\delta\ll\kappa$, according to Eq.~(\ref{slave}), the phase diffusion
of the atoms and light are the same, {\em i.e.}\
$\langle\hat{a}^\dagger(\tau)\hat{a}(0)\rangle
\sim\langle\hat{J}^+(\tau)\hat{J}^-(0)\rangle$. We begin by deriving
equations of motion for
$\langle\hat{\sigma}_{A1}^+(\tau)\hat{\sigma}_{B1}^-(0)\rangle$ and
$\langle\hat{\sigma}_{B1}^+(\tau)\hat{\sigma}_{B2}^-(0)\rangle$:
\begin{equation}\label{g11}
\frac{d}{d\tau}
\begin{pmatrix}
\langle\hat{\sigma}_{A1}^+(\tau)\hat{\sigma}_{B1}^-(0)\rangle\\
\langle\hat{\sigma}_{B1}^+(\tau)\hat{\sigma}_{B2}^-(0)\rangle
\end{pmatrix}=
\frac 1 2
\begin{pmatrix}
X & Y \\
Y & X^*
\end{pmatrix}
\begin{pmatrix}
\langle\hat{\sigma}_{A1}^+(\tau)\hat{\sigma}_{B1}^-(0)\rangle\\
\langle\hat{\sigma}_{B1}^+(\tau)\hat{\sigma}_{B2}^-(0)\rangle
\end{pmatrix},
\end{equation}
where
$X=\gamma_c(N-1)\langle\hat{\sigma}_1^z(0)\rangle-\gamma_c-w+i\delta\,,
Y=\gamma_c N\langle\hat{\sigma}_1^z(0)\rangle\,.  $ We have
systematically factorized:
\begin{eqnarray}
  \langle\hat{\sigma}_{1}^z(\tau)\hat{\sigma}_{A1}^
+(\tau)\hat{\sigma}_{B1}^-(0)\rangle
  &\approx&\langle\hat{\sigma}_{1}^z(0)\rangle\langle
\hat{\sigma}_{A1}^+(\tau)
\hat{\sigma}_{B1}^-(0)\rangle\,,
  \nonumber\\
  \langle\hat{\sigma}_{1}^z(\tau)\hat{\sigma}_{B1}^+
(\tau)\hat{\sigma}_{B2}^-(0)
\rangle
  &\approx&\langle\hat{\sigma}_{1}^z(0)\rangle\langle
\hat{\sigma}_{B1}^+(\tau)
\hat{\sigma}_{B2}^-(0)\rangle\,.
\end{eqnarray}
Similarly, one finds that
$\langle\hat{\sigma}_{A1}^+(\tau)\hat{\sigma}_{A2}^-(0)\rangle$ and
$\langle\hat{\sigma}_{B1}^+(\tau)\hat{\sigma}_{A1}^-(0)\rangle$
satisfy the same equation of motion as Eq.~(\ref{g11}). The solution
of this coupled set is straightforward and shows that both
$\langle\hat{\sigma}_{A1}^+(\tau)\hat{\sigma}_{B1}^-(0)\rangle$ and
thus also $\langle\hat{a}^\dagger(\tau)\hat{a}(0)\rangle$ evolve in
proportion to the exponential:
\begin{equation}\label{spectrum}
  \exp\left[-\frac 1 2 \left(w+\gamma_c-(N-1)\gamma_c
      \langle\hat{\sigma}_{1}^z\rangle-\sqrt{(N\gamma_c\langle\hat
        {\sigma}_{1}^z\rangle)^2-\delta^2}\right)\tau\right],
\end{equation}
which we parametrize by $\exp\left[-(\Gamma+i\Delta)\tau/2\right]$,
where $\Gamma$ represents the decay of the first-order correlation and
$\Delta$ the modulation frequency. Laplace transformation yields the
photon spectrum which consists of Lorentzians of halfwidth $\Gamma/2$
centered at frequencies $\pm \Delta/2$.
\begin{figure}[h]
  \centerline{\includegraphics[width=0.75\linewidth, angle=0]{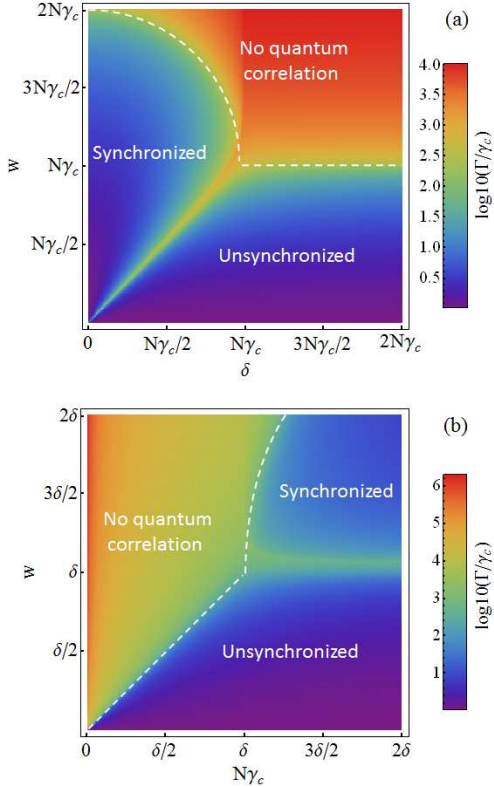}}
  \caption{\label{Fig3}(color online)~(a)~Nonequilibrium phase diagram
    of the quantum synchronization represented by $\Gamma$~(in units
    of $\gamma_c$) on the $w$-$\delta$ parameter plane, where the
    dissipative coupling $N\gamma_c$~($N=10^4$) is fixed. An abrupt
    peak is observed at the boundary between the synchronized and
    unsynchronized phases. (b)~As for (a) but on the $w$-$N\gamma_c$
    parameter plane. }
\end{figure}

The importance of the two-time correlation function is that it
provides direct access to the correlated phase dynamics of the two
ensembles. The parameter $\Delta$ physically represents the precession
frequency of the phase of the collective mesoscopic dipoles with
respect to one another. In Fig.~\ref{Fig2}, we show $\Delta$ as a
function of~$\delta$ at $w=N\gamma_c/2$ for several values of $N$. For
large detuning, $\Delta$ approaches $\delta$, indicating that the
dipoles precess independently at their uncoupled frequency. Below a
critical $\delta$, we find $\Delta$ to be zero, indicating
synchronization and phase locking.

The fact that this system undergoes a synchronization transition that
is fundamentally quantum mechanical and thus quite distinct from the
classical synchronization previously discussed for coupled oscillators
is evident in the observed properties of the linewidth $\Gamma$ of the
Lorenzian peak(s), representing the relative quantum phase diffusion
of the collective dipoles.  This system has three independent control
variables; the detuning $\delta$, the dissipative coupling $N\gamma_c$
and the pumping $w$, so we show $\Gamma$ on the $w$-$\delta$
parameter plane in Fig.~\ref{Fig3}(a) and on the $w$-$N\gamma_c$
parameter plane in Fig.~\ref{Fig3}(b).

In the region of no quantum correlation, the quantum noise due to
pumping destroys the coherences between spins faster than the
collective coupling induced by the cavity field can reestablish
them. Therefore the mesoscopic dipole is destroyed and the observed
spectra are broad.  In both the synchronized and unsynchronized
regions, spins within each ensemble are well-correlated so that the
corresponding Lorenzian peaks have ultranarrow linewidth. As is
apparent in Fig.~\ref{Fig3}(a), the two ensembles cannot be
synchronized when $N\gamma_c<\delta$ since then the coherent coupling
is not sufficient to overcome the relative precession that arises from
the detuning.

For strong coupling, $N\gamma_c>\delta$, the synchronization
transition occurs as the pump rate passes a critical value. The two
phases on either side of the critical region are abruptly
separated. As one approaches the synchronized phase from the
unsynchronized one by variation of either $\delta$ or~$w$, the
linewidth increases rapidly, showing amplification of the effect of
quantum noise in vicinity of the critical point. After passage of the
critical region, the linewidth drops rapidly, leading to rigid phase
locking between the two collective dipoles.

We emphasize that the synchronization dynamics shown in
Fig.~\ref{Fig2} and \ref{Fig3} is a dynamical phase
transition~\cite{Zoller10,Zoller11,Cirac12,Cirac13} that is
reminiscent of a second-order quantum phase transition.
\begin{figure}[t]
  \centerline{\includegraphics[width=0.7\linewidth, angle=0]{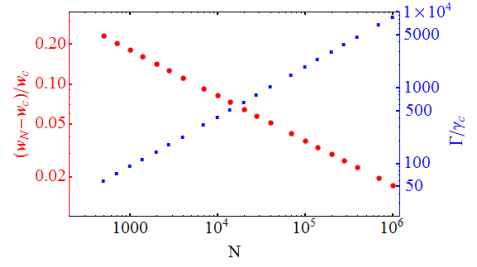}}
  \caption{\label{Fig4}(color online)~Finite size scaling behavior of
    the quantum criticality for $\delta=N\gamma_c /2$. For
    $N\rightarrow\infty$, the critical pump rate is $w_c=\delta$. The
    red dots show the offset between the critical pump rate $w_N$ for
    finite $N$ and $w_c$. The blue squares show $\Gamma$~(in units of
    $\gamma_c$) at~$w_N$. Both exhibit linear scalings on the log-log
    plot.}
\end{figure}
To capture features of the quantum criticality, we numerically study
the finite size scaling behavior.  Fig.~\ref{Fig4} shows both the
critical pump rate $w_N$ for finite $N$ and the corresponding $\Gamma$
at $w_N$. The scaling laws of $(w_N-w_c)/w_c\simeq N^{-0.34}$ and
$\Gamma/\gamma_c\simeq N^{0.66}$ can be identified.

In Hamiltonian systems, a quantum phase transition results from the
competition between two noncommuting Hamiltonian components with
different symmetries on changing their relative weight. The transition
between the two distinct quantum phases can be identified from the
nonanalytical behavior of an order parameter, and the scaling behavior
of certain correlation functions that diverge at the critical point.
By analogy, the synchronization phase transition is caused by the
competition between unitary dynamics that is parametrized by $\delta$
and enters asymmetrically for the two ensembles, and
driven-dissipative dynamics parametrized by $\gamma_c$ that is
symmetric. The order parameter $\Delta$ is zero in the synchronized
phase and non-zero in the unsynchronized phase. The critical behavior
is encapsulated by the divergence of the relative quantum phase
diffusion. It should be emphasized that the treatment given here is
quite different to the typical analysis since the transition is
embodied by the characteristic features of the two-time correlation
functions, rather than the behavior of an energy gap or correlation
length.

In the thermodynamic limit, simple expressions for
$\langle\hat{\sigma}_{1}^z\rangle$ to leading order in $1/N$ can be
obtained:
\begin{equation}\label{simple}
\langle\hat{\sigma}_{1}^z\rangle = \left\{ \begin{array}{rl}
\frac{w}{2N\gamma_c}, &\mbox{ if $\delta=0$} \\
\frac{w^2+\delta^2}{2wN\gamma_c}, &\mbox{ if $0<\delta<w$}\\
\frac{w}{N\gamma_c}, &\mbox{ if $\delta\ge w$}
\end{array}, \right.
\end{equation}
where $w$ should be such that $\langle\hat{\sigma}_{1}^z\rangle<1$.  A
critical point at $w_c=\delta$ can be found by substituting
Eq.~(\ref{simple}) into Eq.~(\ref{spectrum}).  In particular,
$\Delta=(\delta^2-w^2)^{1/2}$ in the unsynchronized phase, which shows
an analogous critical exponent to that of a second-order quantum phase
transition, {\em i.e.}, $\beta=1/2$.

In conclusion, we have presented a system that exhibits quantum
synchronization as a modern analogue of the Huygens experiment but is
implemented using state-of-the-art neutral atom lattice clocks of the
highest precision. It will be intriguing in future work to study the
many possible extensions that are inspired by these results, such as
the effect of an atom number imbalance on the synchronization
dynamics, and the sensitivity of the phase-locking to external
perturbation.

We acknowledge stimulating discussions with J. Cooper, J. G. Restrepo,
D. Meiser, K. Hazzard, and A. M. Rey. This work has been supported by
the DARPA QuASAR program, the NSF, and NIST.


\begin{thebibliography}{99}
\bibitem{book1}S. H. Strogatz, {\em Sync: The Emerging Science of
    Spontaneous Order} (Hyperion, New York, 2003).
\bibitem{Huygens}M. Kapitaniak, K. Czolczynski, P. Perlikowski,
  A. Stefanski, and T. Kapitaniak, Phys. Rep. \textbf{517}, 1 (2012).
\bibitem{book2}A. Pikovsky, M. Rosenblum, and J. Kurths, {\em
    Synchronization: A Universal Concept in Nonlinear Sciences}
  (Cambridge University Press, Cambridge, England, 2001).
\bibitem{book3}S. Bregni, {\em Synchronization of Digital
    Telecommunications Networks} (Wiley, Chichester, 2002).
\bibitem{kuramoto0}Y. kuramoto, {\em Chemical Oscillations, Waves and
    Turbulence} (Courier Dover Publications, 2003).
\bibitem{kuramoto}J. A. Acebr\'{o}n {\em et al}., Rev. Mod. Phys. {\bf
    77}, 137 (2005).
\bibitem{Cross04}M. C. Cross, A. Zumdieck, Ron Lifshitz, and
  J. L. Rogers, Phys. Rev. Lett. {\bf 93}, 224101 (2004).
\bibitem{Milburn12}C. A. Holmes, C. P. Meaney, and G. J. Milburn,
  Phys. Rev. E {\bf 85}, 066203 (2012).
\bibitem{Marquardt11}G. Heinrich, M. Ludwig, J. Qian, B. Kubala, and
  F. Marquardt, Phys. Rev. Lett. {\bf 107}, 043603 (2011).
\bibitem{Jain84}A. K. Jain, K. K. Likharev, J. E. Lukens, and
  J. E. Sauvageau, Phys. Rep. {\bf 109}, 309 (1984).
\bibitem{Wiesenfeld96}K. Wiesenfeld, P. Colet, and S. H. Strogatz,
  Phys. Rev. Lett. {\bf 76}, 404 (1996).
\bibitem{Winfree67}A. T. Winfree, J. Theor. Biol. \textbf{16}, 15 (1967).
\bibitem{Oleg09}O. Kogan, J. L. Rogers, M. C. Cross, and G. Refael,
  Phys. Rev. E \textbf{80}, 036206 (2009).
\bibitem{Zhirov06}O. V. Zhirov and D. L. Shepelyansky,
  Phys. Rev. Lett. \textbf{100}, 014101 (2008).
\bibitem{Zhirov09}O. V. Zhirov and D. L. Shepelyansky, Phys. Rev. B
  {\bf 80}, 014519 (2009).
\bibitem{Orth10}P. P. Orth, D. Roosen, W. Hofstetter, and K. LeHur,
  Phys. Rev. B {\bf 82}, 144423 (2010).
\bibitem{Tony12}T. E. Lee and M. C. Cross, arXiv:1209.0742v1 (2012).

\bibitem{mian12}M. Zhang, G. S. Wiederhecker, S. Manipatruni,
   A. Barnard, P. McEuen, and M. Lipson, Phys.
   Rev. Lett. {\bf 109}, 233906 (2012).
 \bibitem{Mari12}A. Mari, A. Farace, N. Didier, V. Giovannetti, and
   R. Fazio, arXiv:1304.5925v1 (2013).
\bibitem{Hriscu13}A. M. Hriscu and Y. V. Nazarov,
  Phys. Rev. Lett. {\bf 110}, 097002 (2013).
\bibitem{Meiser09}D. Meiser, J. Ye, D. R. Carlson, and M. J. Holland,
  Phys. Rev.  Lett. \textbf{102}, 163601 (2009).
\bibitem{Meiser101}D. Meiser and M. J. Holland, Phys. Rev. A
  \textbf{81}, 033847 (2010); D. Meiser and M. J. Holland, Phys. Rev. A
  \textbf{81}, 063827 (2010).
\bibitem{Thompson12}J. G. Bohnet, Z. Chen, J. M. Weiner, D. Meiser,
  M. J.  Holland, and J. K. Thompson, Nature \textbf{484}, 78 (2012).
\bibitem{Thompson121}J. G. Bohnet, Z. Chen, J. M. Weiner, K. C. Cox,
  and J. K. Thompson, Phys. Rev. Lett. {\bf 109}, 253602 (2012).
\bibitem{semi} We have validated the closed set of
  Eq.~(\ref{eqz})--Eq.~(\ref{eqy}) by comparison with exact solutions
  of the quantum master equation based on applying the SU(4) group
  theory~(see Minghui Xu, D. A. Tieri, M. J. Holland, Phys. Rev. A
  \textbf{87}, 062101 (2013)). Due to the presence of multiple
  ensembles it is difficult to implement exact calculations for more
  than about ten atoms.
\bibitem{Zoller10}S. Diehl, A. Tomadin, A. Micheli, R. Fazio, and
  P. Zoller, Phys. Rev. Lett. {\bf 105}, 015702 (2010).
\bibitem{Zoller11}A. Tomadin, S. Diehl, and P. Zoller, Phys. Rev. A
  \textbf{83}, 013611 (2011).
\bibitem{Cirac12}E. M. Kessler, G. Giedke, A. Imamoglu, S. F. Yelin,
  M. D. Lukin, and J. I. Cirac, Phys. Rev. A \textbf{86}, 012116
  (2012).
\bibitem{Cirac13}B. Horstmann, J. I. Cirac, and G. Giedke,
  Phys. Rev. A \textbf{87}, 012108 (2013).
\bibitem{qr}M. Lax, Phys. Rev. \textbf{129}, 2342 (1963);
  C. W. Gardiner, \emph{Quantum Noise} (Springer-Verlag, Berlin,
  1991).
\end{thebibliography}
\end{document}